\documentstyle[epsfig]{mn}
\input{epsf}
\input rotate
\def\beb{\begin{thebibliography}{999}}
\def\eeb{\end{thebibliography}}
\def\bi{\bibitem{}}
\def\etal{{\it et al.}}
\def\deg{^{\circ}}
\newcount\notenumber
\def\clearnotenumber{\notenumber=0}
\def\note{\advance\notenumber by1 \footnote{$^{\the\notenumber}$}}
\clearnotenumber
\bigskip

\title[Pulsar 0943+10's `Drifting'-Subpulse Emission II.] 
{Topology and Polarisation of Subbeams Associated With \\
Pulsar B0943+10's `Drifting'-Subpulse Emission:  \\ 
II. Analysis of Gauribidanur 35-MHz Observations}

\author[Asgekar \& Deshpande] 
{Ashish Asgekar$^{1,2}$~~\& Avinash~A.~Deshpande$^{1}$ \\ 
$^1$Raman Research Institute, Bangalore 560 080 INDIA :
ashish@rri.res.in, desh@rri.res.in\\
$^2$Joint Astronomy Programme, Indian Institute of Science,
Bangalore 560 012 INDIA}
\date{}
\begin{document}
\maketitle

\begin{abstract}
In the previous paper of this series Deshpande \& Rankin (2001)
reported results regarding subpulse-drift phenomenon
in pulsar B0943+10 at 430~MHz and 111~MHz.
This study has led to the identification of
a stable system of subbeams circulating around the magnetic axis of this
star. Here, we present a single-pulse analysis of
our observations of this pulsar at 35~MHz. The fluctuation properties
seen at this low frequency, as well as our independent estimates
of the number of subbeams required and their circulation time, agree
remarkably well with the reported behavior at higher frequencies. We use
the `cartographic'-transform mapping technique developed 
in \mbox{Paper-I} to study the emission pattern in the polar
region of this pulsar. The significance of our results in the context of
radio emission mechanisms is also discussed.
\end{abstract}

\begin{keywords}
pulsars: general ---
pulsars: individual: B0943+10 ---
radiation mechanism: non-thermal
\end{keywords}

\section{Introduction}

The foregoing paper by Deshpande \& Rankin (2000; hereafter
\mbox{Paper-I}) 
reported detailed studies of the steady subpulse drift of pulsar 
B0943+10, using Arecibo polarimetric observations at 430 and 111~MHz.
They introduced several new techniques of fluctuation-spectral 
analysis and addressed the possible aliasing of the various 
features in an effort to determine their true frequencies. The 
paper further introduced a `cartographic' transformation (and 
its inverse) which relates the observed pulse sequence to a 
rotating pattern of emission above the pulsar polar cap. This 
approach provided a direct means of determining some of the key 
parameters characterizing this underlying source pattern and its 
stability---thus facilitating a fuller comparison with the 
predictions of pulsar emission theories (see also Deshpande \& 
Rankin,~1999, hereafter DRa).

While the Arecibo observations in \mbox{Paper-I} document a great
deal about 0943+10's remarkable `drift' behavior, they also 
demonstrate the need for further studies of different types and 
at various wavelengths.  The 430-MHz observations, for instance, 
sample only the outer radial periphery of the subbeams defining 
the overall hollow emission cone of the pulsar---and this is apparently 
a major reason for its unusually steep spectrum above 100 MHz.  
Observations at decameter wavelengths are particularly interesting,
both because the pulsar is expected to be relatively brighter and 
also because they provide a means of exploring the character of 
the larger low frequency conal beam size [often attributed to 
`radius-to-frequency mapping' or `RFM'; {\it e.g.} see 
Cordes,~1978, and the generally assumed dipolar field geometry].
At 35~MHz, then, we would expect our sightline to sample the 
region of emission more centrally, with the result that, the full radial
extent of the subbeams would be `seen' and the high frequency
`single' profile would be well-resolved into a `double' form.  
The character of the low frequency emission pattern would also be 
of interest---both in its own right and as compared to those at 
higher frequencies---for what insights it might provide into the 
manner in which emission sites at different heights/frequencies
are connected.  

Below, we report the results of our analysis of
pulse-sequence observations of B0943+10 at 35~MHz.
Our observations and analysis procedures are outlined in
the next section.  Subsequent sections then describe the results of 
fluctuation-spectral analysis, the implied subbeam-system topology, and 
the derived polar emission maps [some preliminary results 
were reported earlier in Asgekar \& Deshpande~(2000)].

\section{Observations and Analysis}

Observations of this pulsar were carried out as a part of a new programme
of pulsar observations (Asgekar \& Deshpande, 1999) initiated
at the Gauribidanur Radio Observatory (Deshpande, Shevgaonkar \& Shastry, 1989)
with a new data acquisition system (Deshpande, Ramkumar \& Chandrasekaran,
2000).  The pulsar was observed for a typical duration of $\ga{1000}$~s in
each session. Several such observations were made during the spring of 
1999. We discuss two data sets below, denoted as `C' and `G', which
were observed on February 7 and 13, respectively.

From the signal voltages (over a 1-MHz bandwidth) sampled at the Nyquist
rate with 2-bit, 4-level
quantization, a time-series matrix of intensity with a desired spectral and
temporal resolution (equivalent to data from a spectrometer) was obtained.
After checking for, and excising, interference in the spectral as well as the
time domains,
the different spectral channel data were combined after appropriate 
dispersion-delay correction. 
The resulting time sequence was resampled at the desired intervals
of the rotation phase of the pulsar to obtain a gated pulse sequence for
a specified gate-width (typically~$\sim 100\deg$). 

Our techniques of spectral analysis for the 35-MHz observations 
are virtually identical to those used in \mbox{Paper-I}, so we will rely on 
its fuller descriptions of these methods and 
request that the reader refer to them there.

\subsection*{Fluctuation Spectra and Aliasing}

Our 35-MHz observations lack the sensitivity to detect 
a majority of the individual pulses in a sequence, let 
alone to identify their drift pattern.  Any such drift 
can best be investigated using longitude-resolved
fluctuation (hereafter, `LRF') spectra (Backer, 1973). 
\mbox{Figure 1} then gives such a spectrum for the G sequence.
Note in particular 
the prominent peaks at 0.459\,c/$P_{1}$ (feature 1) and 
0.027\,c/$P_{1}$ (\mbox{feature 2}). \mbox{Feature 1} is associated 
with the observed subpulse drift at meter wavelengths;
it is barely resolved here, implying a Q of $\ge$500. This 
clearly indicates that the underlying modulation process is also 
quite steady at decameter wavelengths.
The modulation frequency is slightly different for the C 
observation, but small variations ($\le 0.5\%$) were also 
noted by Deshpande \& Rankin (\mbox{Paper-I}), so are not surprising.
The important questions
we wish to address are the effect of a possible aliasing of 
feature 1 and its relation to feature 2.

\begin{figure}
\epsfig{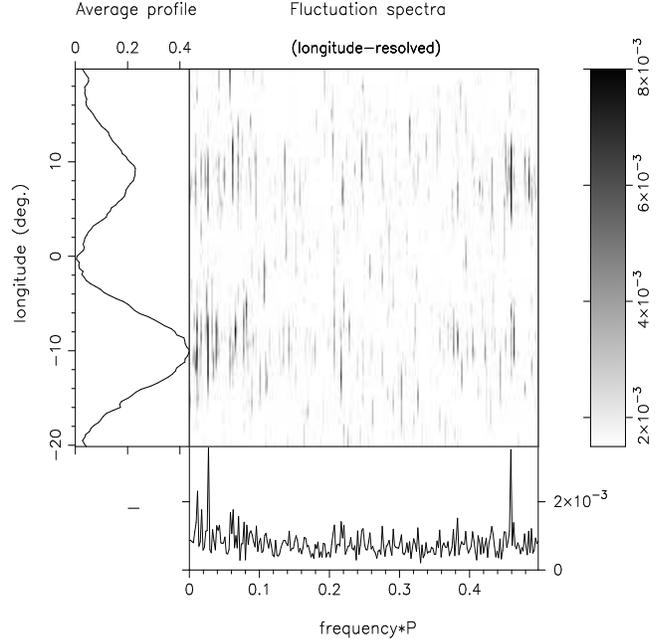}
\caption[]{LRF spectrum of the 34.5-MHz G sequence, 
using 512 pulses. The body of the figure gives the amplitude
of the features in (arbitrary) grey-scale displayed on the right.
The average profile of the pulsar is shown in the left-hand
panel, and the integral spectrum is given at the bottom of the
figure (see \mbox{Paper-I}:\mbox{Figure 1} for details).}
\end{figure}
\begin{figure}
\epsfig{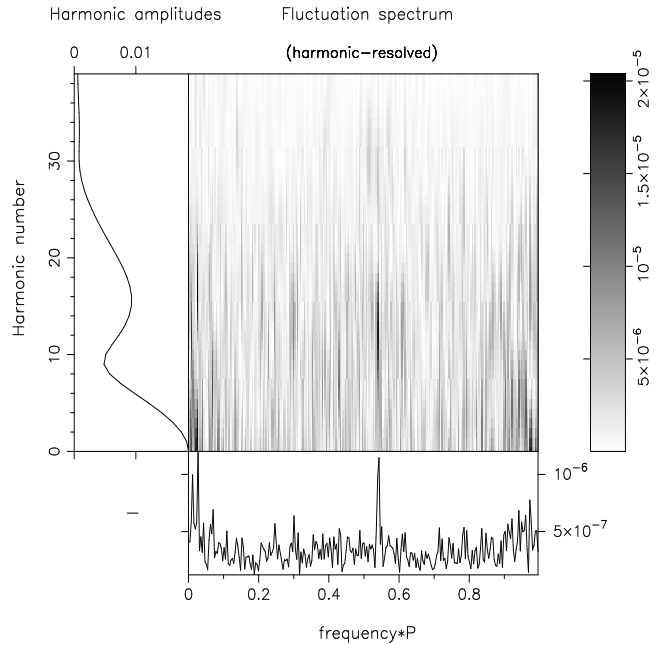}
\caption[]{HRF spectrum of the 34.5-MHz G sequence. The amplitude
of the frequency components at the fundamental rotation
frequency~(1/P$_1$) and its harmonics are plotted in the left-hand
panel of the diagram. The body of the figure gives the amplitude
of the other frequency components in the spectrum upto~40/P$_1$ in
(arbitrary) grey-scale, displayed on the right.The bottom panel
shows the sum of the frequency components, collapsed onto a
1/P$_1$ interval. See \mbox{Paper-I}: \mbox{Figure 4} for details.}
\end{figure}

\begin{figure}
\epsfig{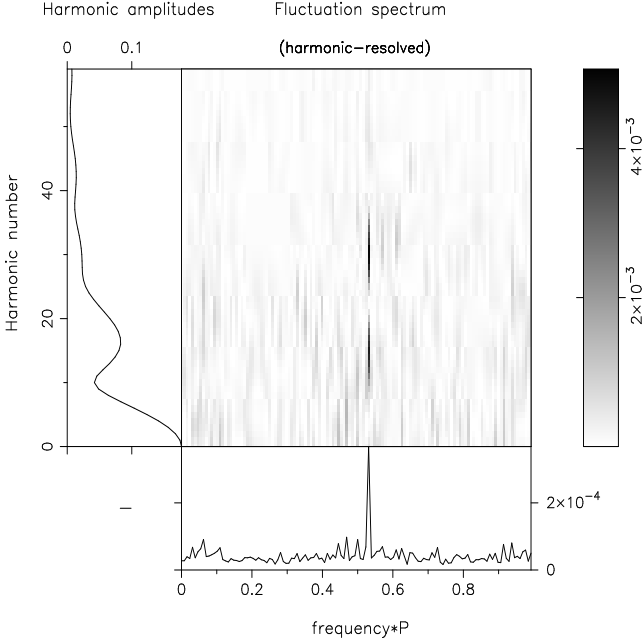}
\caption[]{HRF spectrum of the 34.5-MHz C sequence using 128~pulses.
The \mbox{feature 2} (seen in G~sequence) is not detectable here.}
\end{figure}

To assess the issue of aliasing, we compute a harmonic-resolved 
fluctuation spectrum (hereafter, `HRF' spectrum; see \mbox{Paper-I}:
\mbox{Figure 4}) by Fourier transforming the continuous time series.
Such a time series is reconstructed from the (gated) pulse sequence
by setting the intensity in the off-pulse region to zero. A
spectrum constructed in this manner for the G observation 
is displayed in \mbox{Figure 2}. Note that amplitude modulation would 
produce a pair of features in the HRF spectrum, symmetric about 
frequency 0.5\,c/$P_1$ and with nearly equal amplitudes; whereas, phase 
modulation would produce a strong asymmetry in the pair amplitudes, 
as exemplified in \mbox{Figure 2}.  We can then see that the fluctuation 
feature at 0.541\,c/$P_1$ is associated with a phase modulation,  
and further appears as the first-order aliased feature 1 at
0.459\,c/$P_1$ in the LRF spectrum of \mbox{Figure 1}.  A similar 
conclusion regarding the nature of the modulation can be drawn
by looking at the rate of change of the modulation phase with 
longitude (see \mbox{Figure 5}), which is $26\deg/\deg$---too 
large for a mere amplitude modulation.

We now fold the pulse sequence at the modulation period $P_3$
in \mbox{Figure 4}. The drift-bands related to the two components 
are clearly visible and exhibit a phase difference between 
the modulation peaks.  The `bridge' region between the two 
average-profile components exhibits little discernible 
fluctuation power, so we could be missing one or more drift 
bands in this longitude interval. We therefore express the 
modulation phase difference between the two components as,
$\Delta{\theta} = 2\pi\,(0.8 + m)$ (where, m\,=\,0,1,2..). If 
the modulation under the two components is caused by `passage' 
of the same emission entity through our sightline as it 
rotates around the magnetic axis, the average profile-component 
separation in longitude of  about 20$\deg$ suggests a
$P_2$ value of about $11\deg$ for $m=1$---which is consistent
with the estimates at meter wavelengths. Further, the azimuth
angle subtended by the profile components at the magnetic axis
can be computed using our knowledge of the viewing 
geometry [referring to eq.~(3) of \mbox{Paper-I}], and is estimated to
be about $33\deg$. We adopt here the values of $\alpha$ and
$\beta$, determined in \mbox{Paper-I}---that is, 11.64$\deg$ and
-4.31$\deg$, respectively. It follows that the magnetic
azimuthal separation between the subbeams would be
about~$18\deg$ (corresponding to $P_2$),
suggesting a rotating system of about 20 subbeams. 

It is clear that the 35-MHz observations exhibit phase 
modulation very similar to that noted at higher frequencies, 
and the true modulation frequency is near 0.541\,c/$P_1$, 
implying that $P_3 = 1.848 {P_1}$. This means that the 
subpulse drift proceeds in the direction opposite as that 
of star's rotation. Some enhancement in the spectral power 
is evident at 0.06\,c/P$_1$, corresponding to the aliased 
second harmonic of the primary modulation feature. The 
frequency and Q-value of the fluctuation features are in
excellent agreement with those based on analysis of 430-MHz 
observations.  

The \mbox{feature 2}, at about 0.027\,c/P$_1$, is new and
also has a high~Q. As both LRF- and HRF-spectra show
the feature at the same frequency, we conclude that it is
not aliased and that 0.027\,c/P$_1$ is its true frequency.
It is important to note that this feature, seen in set G,
is {\it not} apparent in the corresponding spectra 
of 111 and 430-MHz data (\mbox{Paper-I}), as well as in our other data
sets at 35 MHz. We discuss its significance below.

\subsection*{The Amplitude-Modulation Feature, Subpulse Spacing
and the Number of Rotating Subbeams}

With an understanding of the intrinsic frequencies of the 
two principal features in the LRF spectrum, we can now 
explore a possible relationship between them. A straightforward 
calculation, $0.541/0.027={20}$, shows that the two features 
are harmonically related within their errors. We have 
looked for the sidebands \mbox{(at~$\pm$ 0.027\,c/P$_1$)} around
the phase modulation feature at~0.541\,c/$P_1$. Such side-bands
were identified in some sections of the 430-MHz sequences
(see \mbox{Paper-I} and DRa), but they occur in our data only
at a much lower level of significance 
({\it e.g.}, see the central panel of \mbox{Figure 1}). The
0.027\,c/$P_1$ 
feature, just as did the 430-MHz sidebands, suggests a slower 
amplitude modulation, with a periodicity, $\hat{P}_3$, of
37.5\,$P_1$ or just 
20\,$P_3$. The magnetic azimuth angle corresponding to the 
longitude interval $P_2$ would be $\sim 18\deg$, if a total of
20~subbeams were to be distributed more-or-less evenly around 
the magnetic axis, a value which is in excellent agreement with 
our estimate above, implying consistency with the viewing geometry 
as determined in \mbox{Paper-I}.

All these circumstances, along with the high $Q$s of the modulation 
features and their harmonic relationship, imply a stable 
underlying pattern of 20 subbeams, with the period of the low 
frequency feature (feature 2) providing a direct estimate of the 
circulation time of this pattern. This slow modulation may also 
be partially a phase modulation, considering the asymmetry 
in the features at 0.027\,c/$P_1$ and 0.973\,c/$P_1$ (see
\mbox{Figure 2}). This would imply some non-uniformity in the spacing of 
the subbeams.

\begin{figure}
\epsfig{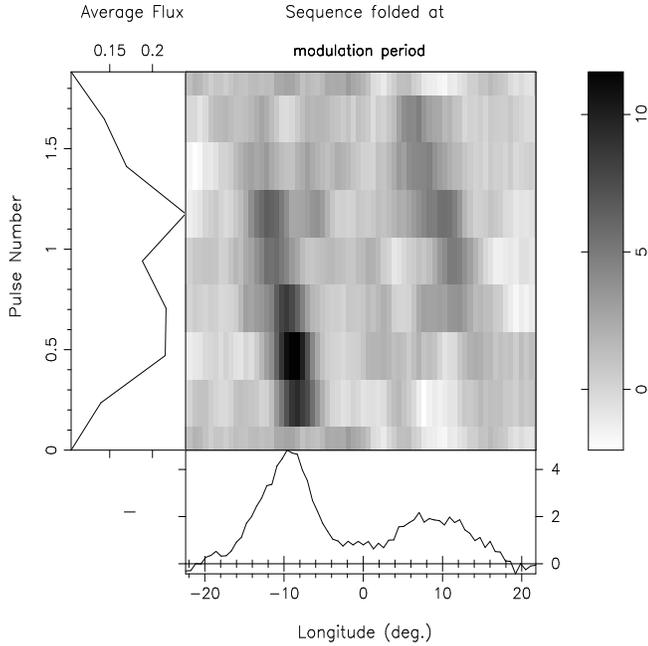}
\caption[]{Diagrams showing the results of folding the C 
observation at the primary modulation period $P_3$. 
Note the separate `drift' bands for two components in 
the central panel. The average profile is plotted in 
the bottom panel, and the average flux at a particular
phase of modulation in the left-hand panel. Note that 
the `drift' band for the trailing component peaks 
earlier than that for the leading component.}
\end{figure}
\begin{figure}
\epsfig{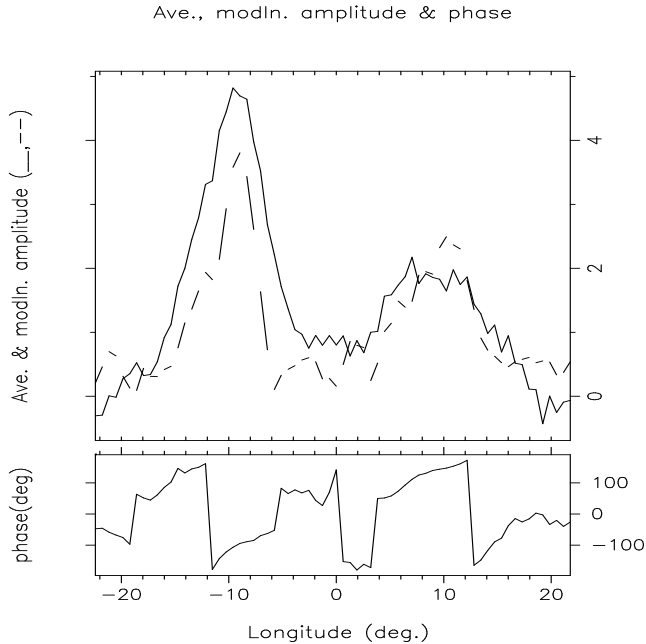}
\caption[]{The phase of the primary modulation $P_3$ as 
a function of pulse longitude for the C observation, 
along with the average (solid curve) and modulated-intensity 
(dashed curve) profiles.}
\end{figure}

\mbox{Figures 6} and~7 show the average polar emission patterns 
corresponding to the G and C observations, respectively.  
These patterns were reconstructed using the `cartographic 
transformation' technique developed in \mbox{Paper-I} as well as 
the various parameters describing our viewing geometry of 
the pulsar determined there, but take all values of the 
circulation time ($\hat{P}_3$) from the present analysis.  
In assembling these maps, we make no use of the subbeam number 
once the circulation time has been determined directly; 
rather, the `cartographic transformation' transforms the 
pulse sequence into the frame of the pulsar polar cap
[see \mbox{Paper-I} : eqs.~5-8], so that the elements of the emitting 
pattern and their distribution can be viewed directly. In
addition, the inverse `cartographic' transform provides a 
useful means of assessing the consistency and uniqueness 
of the input parameters (see Deshpande, 1999 for details).
In spite of our overall poor signal-to-noise observations,
these `closure' consistency checks show a remarkable 
sensitivity to the circulation time, the sign of $\beta$, 
and to the drift direction. Relative insensitivity to the 
other parameters, however, does not allow us to assess 
the viewing geometry independently.

\subsection*{Polar Emission Patterns at 35 MHz}

\mbox{Figure 6} shows the average emission pattern mapped using 
512 pulses from the G sequence. Remarkably, the `active'
subbeams seem to occupy only 
about half of the conal periphery. Such a deep and
systematic modulation in the subbeam intensity around 
the emission cone is completely consistent with the 
appearance of the slow modulation feature in the fluctuation
spectra (see \mbox{Figures 1} and~2). The emission 
pattern of the C observation in \mbox{Figure 7}, on the other
hand, exhibits most of the 20 subbeams and a more-or-less 
uniform spacing---that is, an unsystematic pattern of 
intensities---a picture which is also entirely compatible with
the observed absence of any significant slow modulation 
feature in its fluctuation spectra (\mbox{Figure 3}). It is worth
mentioning that the power associated 
with the primary modulation ({\it i.e.} drift) is 
significantly higher in the C sequence than in the 
G observations.  

\begin{figure}
\epsfig{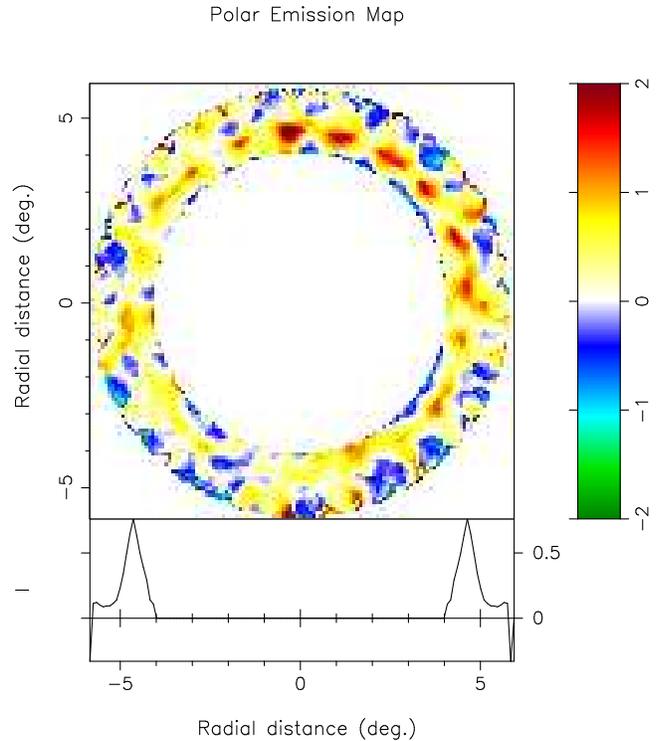}
\caption[]{Average polar emission map made with 512 pulses from the 
G observation.  Note the strong asymmetry about the magnetic axis. See text 
for details.}
\end{figure}

\begin{figure}
\epsfig{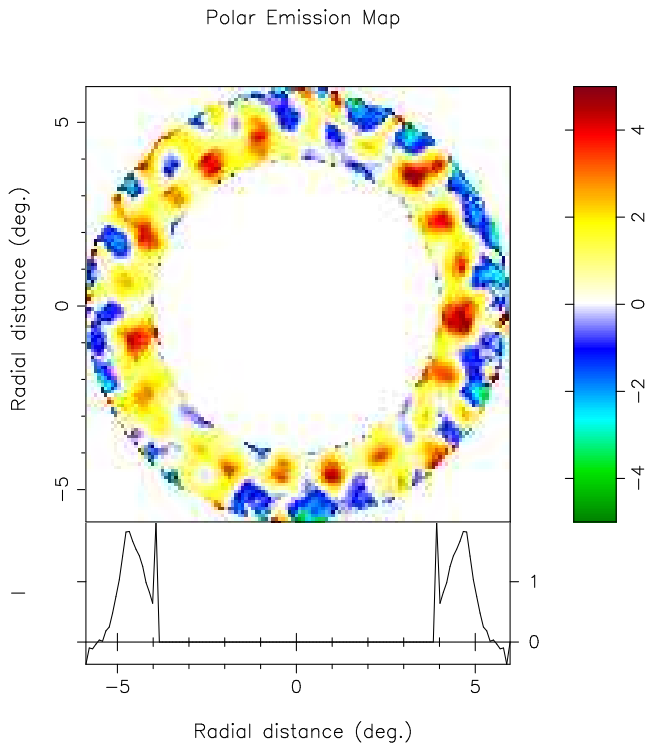}
\caption[]{Average polar emission map made with 140 pulses from the 
C observation. Here we see a nearly full complement of subbeams,
distributed evenly about the magnetic axis.  See text for details.}
\end{figure}

It is clear from both the images that the sampling of the subbeams in
the radial direction is more or less complete. This is significantly so in
contrast with the partial sampling of the subbeams at 430~MHz. The situation
is easily understood in terms of the larger angular radius of the 
radiation cone at low radio frequencies, which permits our fixed sightline 
to sample it more centrally.   The slight elongation (in the azimuthal 
direction) apparent in the subbeam shape may not be real, perhaps resulting 
simply from slight jitter in the positions of the subbeams relative to the 
average.

A series of polar maps, each from successive subsequences 
(of duration $\ge\hat{P}_3$), were generated and viewed in a movie-like 
fashion\footnote{These moving images can be viewed at: 
`http://www.rri.res.in/$\sim$GBD/maps/maps.html'}
Although individual
frames have a reduced signal-to-noise ratio, certain significant 
characteristics can nonetheless be readily discerned. While some
features are stable over only a few circulation times, most ({\it e.g.}, 
the bright subbeams of the G sequence in \mbox{Figure 6}) remain relatively 
prominent over many hundreds of pulse periods. In both observations, 
in a set of successive frames of a movie, a few subbeams usually 
dominate others in intensity. Overall, the brightness of the 
subbeams was observed to fluctuate by up to about a factor of 4 
over the length of the sequence.

In the G sequence a few subbeams are persistently bright, accounting 
for most of the power in the average profile.  A very different
behavior is seen in the C observation, where the subbeams are all 
very weak and unstable in the initial part of the sequence. Then, 
they gradually increase in intensity until most of the 20 subbeams 
constitute a more-or-less stable configuration---that depicted in 
\mbox{Figure 7}---for some four or five circulation times, whereupon they 
again gradually fade back into a disorganized pattern.  The phase 
modulation feature is not readily detectable in either the 
pre- or post-bright 
phases. Neither, interestingly, are associated changes observed in the
shape of the average profile; rather, it maintains a clear `B'-mode 
profile form, apart from the changes in the average intensity.

\section{Discussion}

In the foregoing paragraphs, we have outlined our analysis of 
the fluctuation properties apparent in the decametric emission
of pulsar B0943+10, as well as our efforts to understand the
underlying emission pattern responsible for them.

Although the signal-to-noise ratio of our 35-MHz 
observations is insufficient to detect most individual subpulses, 
we have demonstrated that the average fluctuation behavior can 
nonetheless be reliably characterized, because both the magnitude 
of the fluctuations is large and their pattern relatively stable.
Indeed, such long sequences of pulses present an advantage (and 
opportunity) only if the underlying modulation is relatively stable 
over the relevant time scales, as in the present case. The primary 
phase-modulation frequency is observed to vary, but only on scales 
of many hundreds of pulse periods. Although the observed variations in the
apparent $P_3$ might suggest corresponding fractional changes
in the circulation time, they are more likely to be a manifestation of 
tiny (an order of magnitude smaller) fractional changes 
in the subbeam spacing.

Pulsar B0943+10 is unusual in that conal narrowing at frequencies 
above about 300 MHz causes its beam to progressively miss our sightline.
For a given $\beta$, we expect our sightline to sample the
radiation cone at different $\beta /\rho$ values, such that more  
central traverses occur at lower frequencies---the `single' profile
form of the pulsar at 430 MHz thus evolving smoothly into a well
defined `double' form as $\beta /\rho$ decreases sufficiently at longer 
wavelengths. The average profile of PSR\,0943+10 at 35 MHz exemplifies
this situation with an estimated $\beta /\rho$ of 0.78 (see Table~II
of \mbox{Paper-I}), corresponding to a nearly
complete sampling of the radial `thickness' ($\Delta \rho$) of the 
hollow-conical emission pattern. The fractional thickness of this 
conal beam $\Delta\rho/\rho$ is estimated to be about~$15\%$, 
consistent with that estimated by Mitra \& Deshpande~(1999).

The average profile at decameter wavelength also shows asymmetry 
(the peak intensities of the two components around the apparent
central longitude differ by a factor of about 2), so also the
profile of modulated power (see \mbox{Figure 5}). Such an asymmetry,
though striking, may be traceable 
to relatively small departures of the polar
cap boundary from circularity, if the departures are comparable to the
fractional thickness of the emission cone ($\Delta\rho/\rho$).
A study, by Arendt \& Eilek (2000), of the detailed shape of the
polar-cap boundary and its dependence on the inclination
angle~$\alpha$, shows that an about 5\% departure from circularity is
expected for $\alpha = 15\deg$. Our results, together with those 
of \mbox{Paper-I}, should allow us to estimate the fractional
non-circularity which would cause the observed asymmetries, and 
our efforts in this regard will be reported elsewhere.

The main results of our decametric study of PSR~0943+10's 
fluctuation properties are as follows:
\begin{itemize}
\item We have shown that fluctuation spectra at decameter wavelengths 
also display a narrow feature with a high $Q$, which is related to the 
`drifting' character of its subpulses.
\item We have independently resolved how this feature is aliased and 
have determined the true frequency of the primary phase modulation.
\item The high and low frequency features in the G sequence were found 
to be related harmonically (the former just 20 times the latter). 
Furthermore, 
weak sidebands appear to be associated with the primary phase-modulation 
feature, again suggesting that a slower fluctuation, of period 
\mbox{$\hat{P}_3 = 20 P_3$}, is modulating the primary phase fluctuation.
\item Since the viewing geometry at 35 MHz is too central
($\beta/\rho = 0.78$) to follow the `drift' band continuously 
across the pulse window, we have adopted different means of estimating
the value of $P_2$, which at $11\deg$ 
agrees well with that determined in \mbox{Paper-I}.
\item Our observations lead to the conclusion that a system of
20 subbeams, rotating around the magnetic axis of the star with 
a circulation time $\hat{P}_3$ of about 37 $P_1$, is responsible 
for the observed stable subpulse-modulation behavior.
\item Compared with the higher frequency maps, the 35-MHz maps 
sample the subbeams much more completely.  However, the 35-MHz 
subbeams, on the whole, show much less uniformity in their 
positions and intensities.  
\item At both 35 MHz and the higher frequencies, the intensities 
of individual subbeams fluctuate, maintaining a stable brightness 
only for a few circulation times.  
The stability time-scales of the entire pattern
appear roughly comparable between 35~MHz and 430~MHz. 
\end{itemize}

Since the particle plasma responsible for the pulsar emission originates
close to `acceleration' zone (which is considered to be located
near the magnetic polar cap), whereas the conversion to radio 
emission occurs at a height of several hundred kilometers above 
the stellar surface, we are led to conclude that the polar cap is 
connected to the region of emission via a set of `emission columns'. 
This picture is further supported by the very similar emission 
patterns observed at 35~MHz and 430~MHz---emitted, presumably, at 
significantly different altitudes in the star's magnetosphere.
If indeed some `seed' activity results in the common character of 
the emission patterns at different frequencies, then the parameters
describing the fluctuations ({\it e.g.}~$P_2, P_3$) should be 
independent of the observing frequency, e.g. as in the model of
Ruderman \& Sutherland (1975). This expectation is consistent
with our observations of B0943+10, both here and in \mbox{Paper-I},
as well as with the findings of Nowakowski \etal\ (1982).

We expect this basic picture of subbeams and their apparent
circulation to be valid in other pulsars too, although the
viewing geometry and other quantitative details may differ.
Whether we observe an amplitude or phase modulation
would depend largely on the viewing geometry. For example,
at frequencies much lower than 35~MHz, we expect the phase
modulation to be less discernible in B0943+10,
and the fluctuation would be describable as primarily an
amplitude modulation within each of the subbeam passages
across our sightline. We suspect that the situation at
35~MHz may not be far from such a transition, and therefore
any trivial interpretation of the `apparent' drift-band
separation as P$_2$ would be completely misleading.
Clearly, a different approach is then needed to determine $P_2$ in such a
case, as was done with our data, or for pulsar \mbox{B0834+06}
(Asgekar \& Deshpande, 2000).

Among a total of 6 detections at 35~MHz, we find only one
sequence exhibiting a `Q'-mode behavior---that is, with 
both the average profile and fluctuation spectra 
displaying clear deviations from `B'-mode properties.
This predominant `B'-mode character of the decametric
emission in our observations, as well as in the earlier 
observations of Deshpande \& Radhakrishnan (1994), appears 
inconsistent with the trend at higher frequencies where 
both the `B' and `Q' modes may occur with similar frequency 
(Suleymanova \etal\, 1998). It is possible however, that 
the intervals of disorganized pattern in our observations 
correspond to some mode-change activity. The weakening of 
the pulse after the bright phase in the C~sequence may be 
regarded as such an example.

To summarize, our observations of \mbox{PSR 0943+10} at
35~MHz show 
remarkable similarity with the high frequency observations 
in \mbox{Paper-I} in terms both of their fluctuation properties
and of the underlying emission patterns they represent. 
This similarity, combined with the `radius-to-frequency
mapping', implies that the regions of radio emission
correspond to a well-organized system of plasma columns
in {\it apparent} circulation around the magnetic axis of
star. We emphasis that the observed circulation should not
be mistaken for an unlikely physical (transverse) motion
of the plasma columns in the presence of strong magnetic
fields. The circumstances suggest that a common `seed'
activity is responsible for the generation and motion of
the relativistic plasma.

\section*{Acknowledgments}
We thank the staff at the Gauribidanur Radio Observatory,
in particular \mbox{H. A. Ashwathappa}, \mbox{C. Nanje Gowda},
and \mbox{G. N. Rajasekhar}, for their invaluable help during
observations. We are thankful to V.~Radhakrishnan, Rajaram
Nityananda, and Joanna Rankin for many fruitful discussions
and their comments on the manuscript. This research has made
use of NASA's ADS Abstract Service.

\beb
\bi Arendt P. A. Jr., Eilek, J. A., 2000, ApJ, preprint
\bi Asgekar A., Deshpande, A.~A., 1999, BASI, 302, 27
\bi Asgekar A., Deshpande A.~A., 2000, ASP Conference Series, Vol. 202;
Proc. of IAU Colloquium \# 177, (San Francisco: ASP), Ed. M. Kramer,
N. Wex, and R. Wielebinski, pp. 161
\bi Backer D. C., 1973, ApJ, 182, 245
\bi Cordes J. M., 1978, ApJ, 222, 1006
\bi Deshpande A. A., 2000, ASP Conference Series, Vol. 202;
Proc. of IAU Colloquium \# 177, (San Francisco: ASP), Ed. M. Kramer,
N. Wex, and R. Wielebinski, pp. 149
\bi Deshpande A. A., Radhakrishnan V., 1994, JAA, 15, 329
\bi Deshpande A. A., Ramkumar P. S., Chandrasekaran S., 2000, in preparation. 
\bi Deshpande A. A., Rankin J.M., 1999, ApJ, 524, 1008 ~{\bf DRa} 
\bi Deshpande A. A., Rankin J.M., 2001, MNRAS, 322, 438~{\bf \mbox{paper-I}} 
\bi Deshpande A. A., Shevgaonkar R. K., Shastry Ch. V., 1989, JIETE, 35(6), 342
\bi Mitra D., Deshpande A. A., 1999, A\&A, 346, 906
\bi Nowakowski L., Usowicz J., Kepa A., Wolszczan A., 1982,  A\&A, 116, 158
\bi Rankin J. M., Deshpande A. A., 2000, ASP Conference Series, Vol. 202;
Proc. of IAU Colloquium \# 177, (San Francisco: ASP), Ed. M. Kramer,
N. Wex, and R. Wielebinski, pp. 155
\bi Ruderman M. A., Sutherland P. G., 1975, ApJ, 196, 51
\bi Suleymanova S. A., Izvekova V. A., Rankin J. M., Rathnasree N., 1998, JAA, 19, 1
\eeb
\end{document}